\documentclass[reprint,amsmath,amssymb,aps,prb,showpacs]{revtex4-1}
\usepackage{epsfig}
\usepackage{amsmath}
\usepackage{amssymb}
\usepackage{graphicx}
\usepackage{dcolumn}
\usepackage{hyperref}
\usepackage{bm}
\usepackage{CJK}

\begin{document}

\preprint{APS/123-QED}
\begin{CJK*}{GB}{}
\CJKfamily{gbsn}
\title{Theoretical Description of Scanning Tunneling Potentiometry}
\author{Weigang Wang(Íõκ¸Õ)}
 \email{weigwang@stanford.edu}
\author{Malcolm R. Beasley}
\affiliation{Geballe Laboratory for Advanced Materials,\\
Stanford University, Stanford, CA 94305
}%
\date{\today}

\begin{abstract}
%To address needs from experiment, we develop a theoretical framework of understanding scanning
%tunneling potentiometry (STP) operation in terms of the density matrix of quantum mechanics. Based on our purpose we do not concern detailed calculation of the density matrix of the sample and leave that for further study. We first provide a
%heuristic understanding of STP in general terms. We then work out the total
%tunneling current based on the density matrix of the sample within the
%general framework of quantum transport. Explicit expressions of measured voltages were obtained under certain assumptions. After
%that, we describe a preliminary setup using local density matrices to handle the
%situation of local mesoscopic physics in an otherwise macroscopic sample, where only sample area near the probe is of concern in the problem. Finally we present some limiting cases.

A theoretical description of scanning tunneling potentoimetry (STP) measurement is presented to address the increasing need for a basis to interpret experiments on macrscopic samples. Based on a heuristic understanding of STP provided to facilitate theoretical understanding, the total tunneling current related to the density matrix of the sample is derived within the general framework of quantum transport.  The measured potentiometric voltage is determined implicitly as the voltage necessary to null the tunneling current. Explicit expressions of measured voltages are presented under certain assumptions, and limiting cases are discussed to connect to previous results. The need to go forward and formulate the theory in terms of a local density matrix is also discussed.
\end{abstract}

\pacs{07.79.-v,72.10.Bg,73.23.-b,73.50.-h}

\maketitle
\end{CJK*}

\section{Introduction}
The use of scanning probes is now widespread.  However, very few of these probes measure transport at very short length scales. The ultimate probe for such measurements is scanning tunneling potentiometry (STP)\cite{STP1st,STP}, in which a scanning tunneling microscope (STM) is used to measure the local potential due the flow of an applied current.  However, only recently have we and others developed STP instruments  that can routinely make measurements under a wide range of conditions and essentially at the fundamental noise limit of STP.\cite{rozler,other1,other3,other2}.

\begin{figure}
\begin{center}
\includegraphics[width=3in,bb=8 8 130 94]{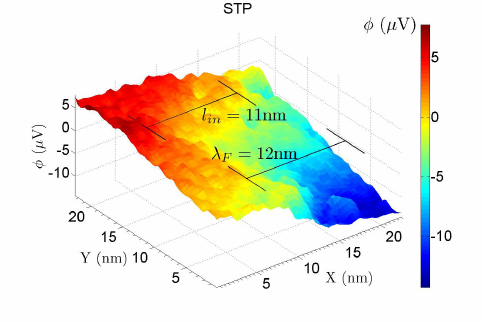}
\caption{STP data taken on epitaxial graphene as an example of transport measurement at length scales smaller than all the length scales relevant in the transport. As seen in the figure, on top of the average gradient, STP shows variations on the nanometer length scale.}\label{STPDemo}
\end{center}
\end{figure}

With our STP, we have focussed on local quantum transport in macroscopic materials, as distinct from  quantum transport through nanostructures (e.g., single molecules, nanotubes and lithographically produced nanostructures).  Our work demonstrates that
potential maps can be obtained in appropriate materials down to distances smaller than all the length scales relevant in transport (the inelastic scattering length, the elastic scattering
length and even the Fermi wave length). An example of an STP image on these very short length scales obtained in our laboratory is shown in Figure \ref{STPDemo} for an epitaxial graphene sample.   As can be seen in the figure, there is considerable local structure in the potential that is relatively large in magnitude.  It is the lack of a quantitative basis for interpreting such images that motivates this paper.  We hope both to explain the theoretical issues for experimentalists and to motivate the needed theoretical extensions for theorists.

There are two issues.  First, it is not clear exactly what potential is being measured.
Macroscopically, the measured potential would be the usual local electrochemical
potential.  However, at such short length scales as STP probes, one cannot use thermodynamic concepts. Second, the situation is inherently quantum mechanical and any calculation of the potential must include the underlying quantum mechanical processes.

These issues are not entirely new.  To some degree they have been addressed in the theories relevant to transport through nanostructures where the sample is small compared to characteristic lengths of the transport.   On the other hand, in the case of these nanostructures, in most of the cases the measurement contacts are relatively macroscopic (compared with an STP tip), and cannot be scanned.  Moreover, the interfaces between the contacts and the sample play an important role in the overall transport.   Numerous theories have been developed to deal with this situation.  Some of them also consider what an STP would measure inside a nanostructure \cite{buttiker1,buttiker2,small1,small2,small3}, although there are no existing STP data yet that can be compared with these theories.  In any event, regarding  the specific question of what STP measures, it was possible to write down explicit expressions for the STP voltage (for example, see equation (37) of reference \onlinecite{buttiker1}). On the other hand, these expressions are explicitly dependent on the voltages applied on the current leads.   By contrast,  in the case where the sample is macroscopically large, the geometry and microscopic processes present in the leads obviously cannot matter.  Hence a proper theoretical formulation is required, and precisely what STP measures remains unclear.

In this paper, we take some first steps toward a theoretical description of a  STP measurement of local quantum transport in a macroscopic sample.  The relevant quantum mechanical quantity is the density matrix, and it is possible to develop an implicit (and under some approximations, an explicit) relation from which the measured potential can be determined in terms of the density matrix and the tunneling process into the sample.  While the tunneling process associated with the STM tip can be accounted for, it is a separate matter how to do the calculation of the density matrix under the relevant nonequilibrium
conditions and short length scales in a macroscopic sample. We do not address these computational challenges in this paper.  As stated above, our goal is to understand scanning tunneling potentiometry in terms accessible to experimentalists and to define the deeper theoretical questions needing attention. Toward these ends, we use both heuristic and more formal approaches.   More specifically, in the formal development, our results pose new problems in quantum transport that need proper theoretical attention in order to have a complete theory suitable for interpreting experiment. One example is the need for formulations in terms of local density matrices as opposed to global ones.

We note that some of the issues discussed above  have been addressed by Chu and Sorbello \cite{sorbello} in the context of their calculation of the residual resistivity dipole of a scattering center in the spirit of Landauer\cite{landauer}. As noted in their paper, the potential measured in STP is not the same as the local electrostatic
potential in the material generated by an electric current. Our work can be thought of as a generalization of their work and an articulation more from the general point of view of a theory of STP measurement. Compared to Chu and Sorbello, our result is more general; and in the final result, instead of a weighted sum, we obtained a form which consists of matrices either being density matrix of the sample or defined with sample and tip wave functions.

%I THINK WE MIGHT JUST DROP THE FOLLOWING PARAGRAPH.  IT REPEATS WHAT WE HAVE ALREADY SAID AND MAY BE TOO MUCH IN AN INTRODUCTION.
%
%Our theory formally fits into the general framework of quantum
%transport\cite{datta}, developed primarily for small samples (first set of problems as discussed above), as a special case. In
%this framework, a density matrix is required to describe the
%sample where our two contacts to the sample are regarded as
%electron reservoirs that are in thermal equilibrium and described
%by Fermi-Dirac distributions with their respective chemical
%potentials. The interaction between the contacts and the sample is
%described by source terms and scattering matrices in the
%nonequilibrium Green's function (NEGF) -- Landauer approach. In
%our theory, the STM tip is a special contact, one with minimal
%source terms under measurement conditions and with specific
%scattering matrices that are related to electron wavefunctions of
%both the tip and the sample.

This paper is organized as follows.
%First, we discuss STP in the
%classical limit by way of introduction and to clarify concepts.
%This discussion also serves to permit emphasis of the general
%feature of potentiometry that what is measured under all
%conditions is the potential necessary to achieve zero current
%though the potentiometric contact. A heuristic approach of
%understanding STP in general terms is presented in this spirit.
First, we describe the setup for STP measurement, also serving to
emphasize that what is operationally measured is the
potential necessary to achieve zero current through the
potentiometric contact. We provide a heuristic understanding of
STP measurement in general terms. Next, we present a simple model
of only two scattering centers with mesoscopic distance to
motivate a more general theory. This leads naturally to the use of
the density matrix of the sample in the expression of the
tunneling current detected in STP and thereby gives meaning to
what is being measured. After
that, we discuss the need to use a local density matrix in the case of macroscopic samples.
Finally, we discuss several limiting cases of our theory to
illustrate better its physical content.

\section{Scanning tunneling potentiometry}\label{stp}
\subsection{STP setup}
\begin{figure}
\begin{center}
\includegraphics[width=3.2in,bb=4 6 139 112]{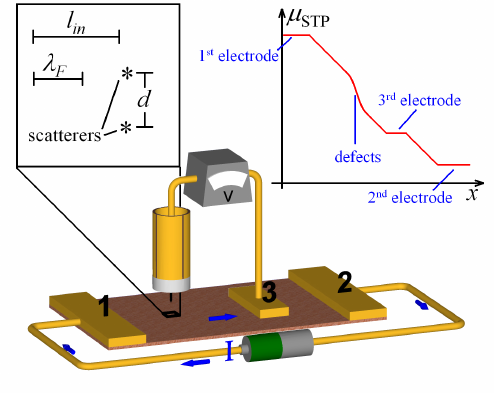}
\caption{Schematic setup of scanning tunneling potentiometry, and model problem to be considered in this paper.
Electrodes 1, 2 and 3 are macroscopic contacts, while the STM tip,
microscopically connected to the sample, functions as the fourth
electrode. The sample
is macroscopic in size and
STP probes a small region which is enlarged to the
upper left. The region has two scattering centers, and the rest
of the sample is assumed to be homogeneous and defect-free, with no
defects (scatterers) present. The distance between the two
defects, the Fermi wavelength and the inelastic mean free path of
the sample are all comparable. To the upper right presents an expected STP result on a large length scale, including the electrodes.}\label{STP3}
\end{center}
\end{figure}

As seen in Figure \ref{STP3}, STP is
effectively a four-point transport measurement using an STM tip. In an STP
measurement, a floating current source provides a current through the sample via
electrodes 1 and 2; a voltage is applied between the third
electrode and the STM tip, which serves as the fourth electrode.
The voltage is so adjusted that the tunneling current between the
sample and the STM tip is zero -- the definition of a potentiometric measurement. This applied voltage which nulls the tunneling current is the data that is recorded in an STP
measurement. Moreover, the capability of STM to scan on nanometer
scales makes STP measurement a nanoscale transport measurement.
The outcome of the experiment is in the form of a potential map of the scanning area accompanied by a topographical map (obtained by conventional STM operation) of the
same area taken point by point successively with the potential  (see Ref \onlinecite{rozler} for details).

\subsection{A heuristic interpretation of STP}

Now consider a macrocopic sample where the current contacts are far removed, specifially where the sample size and the distance between the current contacts are large compared to characteristic transport length scales.  Our goal is to address the question what STP really measures under these conditions. This measured value in STP would correspond to the local electrochemical potential in the thermodynamic limit (i.e., a large distance between the voltage measurements and a large voltage contact area in addition to a macroscopic sample size).   In STP the situation is more subtle.

\begin{figure}
\begin{center}
\includegraphics[width=2.5in,bb=0 0 110 82]{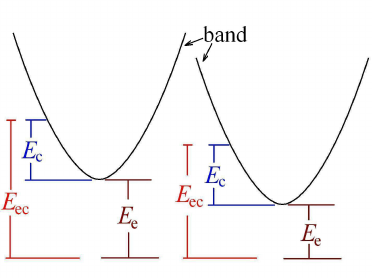}
\caption{Definitions of different energies in the problem.
Chemical energies ($E_c$) are defined with respect to the band,
e.g., the band bottom; electrostatic energies ($E_e$) are energy
differences associated with the band bending due to existence of
an electrostatic field; and the sum of these two energies are the
electrochemical energies ($E_{ec}$). Electrons at different
positions of the sample with the same chemical energy and wave
vector have the same wave function, while they do not necessarily
have the same electrochemical energy due to band bending. Although
electrochemical energy is the appropriate energy one needs to use
when comparing energies at different positions, it is easier to keep track of quantum states with chemical energies.}\label{energies}
\end{center}
\end{figure}

In a conventional STM measurement, the total tunneling current is
written\cite{stmbook}:

\begin{equation}\label{conventional}
I=\frac{4\pi e}{\hbar}\sum_{\vec k, \vec k'}\big[f_s(\vec
k)-f_t(\vec k')\big]\delta(E_{sc,\vec k}-eV_s,E'_{tc,\vec
k'}-eV_t) |M|^2,
\end{equation}
where $\delta(E_1,E_2)$ is the Kronecker delta symbol,
$f_i \ (i=s,\ t)$ are Fermi-Dirac distribution functions of the
sample and the tip, respectively; $|M|^2$ is the magnitude of
tunneling matrix elements, which are assumed to be contant in this equation;
$E_{ic,\vec k}$ are eigen energies of the specific states of the
sample and the tip, where the energies here are measured with respect to the
band (``chemical energy'', see Figure \ref{energies} for clarifying
definitions of chemical energy, electrostatic energy and the sum
of these two, the electrochemical energy); and $-eV_i$ are the energies
associated to the electrostatic potential in the sample and the
tip (``electrostatic energy'', also see Figure \ref{energies}).

%The distribution function of the sample is a Fermi-Dirac
%distribution if there is no current on the sample, however, when
%current do exist, as in the case of STP measurement, the sample is
%not in thermal equilibrium, and the distribution function is not
%expected to be a Fermi-Dirac distribution. One can thus view the
%measurement as probing the non-equilibrium distribution (and
%coherence, as shown in next section) of electrons in the sample by
%changing the electrochemical potential of the tip, and the
%recorded data points are electrochemical potential differences of
%the tip that are necessary to keep the tunneling current zero at
%different positions.

Two modifications need to be made in order to use the form of
equation (\ref{conventional}) in the case of a potentiometric transport measurement. First, the
distribution function of the sample is not one in equilibrium due to
existence of an applied current through the sample, and hence is not a Fermi-Dirac distribution. And second, the magnitude of tunneling matrix elements $|M|^2$
should not be assumed as a constant for each tunneling channel, hence in the
expression one should use an equivalent average value for the
magnitude of tunneling matrix elements.

%From equation (\ref{conventional}) it is seen that the total
%tunneling current in the junction is a weighted sum of the
%difference of distribution functions of the two sides, the weight,
%assuming constant tunneling matrix element magnitudes, is the
%product of density of states of the two sides, reflecting the fact
%that tunneling currents between different channels are incoherent
%and hence should be added to obtain the total tunneling current.

\begin{figure}
\begin{center}
\includegraphics[width=3in,bb=10 0 80 75]{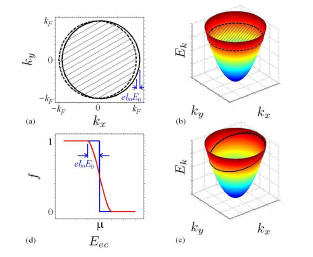}
\caption{Comparison of distribution functions  with and without current. (a) When there is a current, the equilibrium Fermi surface (dashed line and shaded) is shifted to the right (solid line). (b) When viewed in the $E-\vec k$ diagram, in equilibrium, the Fermi surface is a horizontal circle, (c) while when current exists, the Fermi surface is tilted. (d) The occupation rate for each energy, i.e., the distribution function (solid red curve), is different than Fermi-Dirac distribution (blue curve). The distribution function in more complicated situations, for example, when elastic scattering exists, is expected to be different and more interesting than illustrated here.}\label{disfunc}
\end{center}
\end{figure}

Consider now an example of  the distribution function in an non-equilibrium transport situation.  In the linear response region of a homogeneous sample
with no defects and only inelastic scattering with mean free path $l_{in}$, the set of electrons with a certain direction of
wave vector is described by an effective chemical potential
dependent on this direction\cite{sorbello},
i.e., $f_s(\vec k)\propto(e^{\beta(E-\mu_{\hat\theta})}+1)^{-1}$
(here $\beta\equiv1/(k_BT)$, see also Section \ref{limiting}). A distribution function
$f_s(E_{s,ec})$ can be defined as the average probability of
occupancy for each energy $E_{s,ec}$, where the averaging runs
over all degenerate states (with different direction of wave
vector, for example). Plotting $f_s(E_{s,ec})$ as a function of
$E_{s,ec}$, it is expected that $f_s\approx1$ at low energy and
$f_s\approx0$ at high energy, but in the transition region between low and high energies, it is
not a Fermi-Dirac distribution. If we ignore the broadening of the distribution function due to finite temperature, with a constant current
and a cylindrical Fermi surface, in the previous case where the sample is homogeneous and defect-free, the distribution function is in the form
(see also Figure \ref{disfunc} for more detailed clarification):
\begin{widetext}
\begin{equation}\label{heuristic1}
f_s(E_{s,ec})=\left\{\begin{matrix}\\1\qquad & E_{s,ec}<\mu-el_{in}E_0\\
\\
1-\frac{\displaystyle (el_{in}E_0-\mu+E_{s,ec})^{3/2}}{\displaystyle
2(el_{in}E_0)^{3/2}}\qquad &
\mu-el_{in}E_0<E_{s,ec}<\mu\\
\\
 \frac{\displaystyle
(el_{in}E_0+\mu-E_{s,ec})^{3/2}}{\displaystyle 2(el_{in}E_0)^{3/2}}\qquad &
\mu<E_{s,ec}<\mu+el_{in}E_0\\
\\0\qquad &
E_{s,ec}>\mu+el_{in}E_0\\\end{matrix}\right.
\end{equation}
\end{widetext}
where $E_0$ is the electric field and $l_{in}$ is the inelastic mean free path.

Comparing the distribution function of the sample to that of the tip  (a Fermi-Dirac
distribution), it is clear that electrons with relatively low
energies will tend to flow from the tip to the sample, and
electrons with relatively high energies will tend to flow in the
opposite direction. Hence the total tunneling current is a weighted sum
of the difference of the two distribution functions. The weight
for each energy is clearly related to the density of states,
however, this weight is also related to the tunneling matrix
elements as they are determined by wave function values at a
certain position (tip center of curvature) and should not be
assumed to be the same as in equation (\ref{conventional}). One
can write:
\begin{equation}\label{heuristic2}
I=\frac{4\pi
e}{\hbar}\int_{-\infty}^{\infty}\big[f_s(\epsilon)-f_t(\epsilon-eV_t)\big]N_s(\epsilon)N_t(\epsilon-eV_t)\overline{|M|^2}\mbox{d}\epsilon\
,
\end{equation}
where $V_t$, which is the voltage on the tip, effectively moves the
distribution function $f_{t}$ (blue curve
in Figure \ref{disfunc}(d)) in the horizontal direction, such that the total tunneling current equals zero. Here, the densities of states are more or less
semi-classical while the average magnitude of the tunneling matrix
represents the quantum interference, which we will study in detail
in next section. This concept that the total current is
proportional to a weighted sum of the difference of two
distribution functions is not unlike thermal electric effect,
although in that case it is the difference in temperature that is the
origin of the difference in distribution function, while in our
case it is the nonequilibrium nature of current flow.

From equation (\ref{heuristic2}) it is clear that STP does not
measure a well defined thermodynamic potential, rather, it
measures the non-equilibrium distribution function of the sample
with the weight for each energy due to both densities of states
and quantum interference effects in the tunneling matrix. To
actually calculate this weight, a density matrix is needed, which
is described below. It is also clear that even with the same
distribution function, due to relative change in the tunneling
matrix element from quantum interference from one measurment point to another, the STP voltage can change.
This accounts for the STP fluctuation in Chu \& Sorbello's
paper\cite{sorbello}. The basic message of this heuristic consideration is that the STP potential reflects both changes in the distribution functions and the tunneling matrix elements as a function of position.  These insights provide a qualitative basis for considering experiment and set the stage for a more exact theoretical treatment.

\section{General theory in terms of density matrix}
\subsection{Simple model considered in the problem}
%\begin{figure}
%\begin{center}
%\includegraphics[width=3in,bb=0 0 130 120]{modelproblem2.eps}
%\caption{Model problem to be considered in this paper. The sample
%is macroscopic in size and
%STP probes a small region which is enlarged to the
%upper left. The region has two scattering centers, and the rest
%of the sample is assumed to be homogeneous and pristine, with no
%defects (scatterers) present. The distance between the two
%defects, the Fermi wavelength and the inelastic mean free path of
%the sample are all comparable. To the upper right presents an expected STP result on a large length scale, including the electrodes.}\label{model}
%\end{center}
%\end{figure}

%In our STP measurement on epitaxial graphene (reported
%elsewhere\cite{wang}, also see Figure \ref{STPDemo} for an example), which has
%characteristic transport lengthscales of nanometer size, we
%demonstrated that STP measurements are possible on these lengthscales. In order to understand STP measurement
%on this lengthscale, one needs to construct a general theory
%describing the measurement.

We introduce a simple model presented in Figure \ref{STP3}, to represent a macroscopic sample with defects closer to one another than the inelastic mean free path. In the
problem, the sample is macroscopic and connected to the current
providing electrodes macroscopically. The STM tip probes a small
region on the sample, and inside this small region are two
scattering centers near each other. The rest of the sample is
assumed to be defect-free and homogeneous, with inelastic mean free
path $l_{in}$ and Fermi wavelength $\lambda_F$. These two
parameters are both comparable to the distance between the two
defects in the problem. The electrochemical potential on the
sample is expected to be a linear curve with respect to position except for where the defects are,
as also shown in the figure. However, as one expects from physical
consideration, and explained in detail below, when comparing STP
data taken at nearby points, such thermodynamic concepts are
inadequate.

In this problem, one can ask the following two questions: a) what
is the STP measurement result near the two defects? b) what is the
cross section of the two defects, if seen far away and treating
them as one single defect? The answer to the second question
determines STP measurement result far away from the defects, as
in that case STP measures the Landauer resistive dipole
potential\cite{landauer}. One needs to
solve a similar problem to answer both questions.

\subsection{Quantum transport approach}
In the quantum transport
approach\cite{datta}, one uses a density
matrix\footnote{Formally, to include all the information one needs
to use correlation functions which also give the phase difference
between states at different time, whereas as we show below, the
density matrix of the sample which only specifies phase difference
between states at the same time is sufficient to describe the
situation.} to describe the sample, and treat all three contacts
as electron reservoirs that have Fermi-Dirac distributions. The
necessity of density matrix in solving the model problem posed
above can be appreciated with the following consideration, with
plane waves as the basis set for electron wave functions:

Suppose we start with an incident plane wave. The plane wave will
be scattered by either of the two defects. The scattered wave is
coherent with the incident wave. As it propagates, the magnitude
of the coherent part shrinks due to inelastic scattering, which
generates waves having indefinite phase difference with the
incident and scattered waves. Moreover, the scattered coherent
wave can be scattered by the other defect, generating a coherent
secondary scattered wave. In short, the incident plane wave will
be scattered back and forth by both scattering centers, with
decaying coherent part. In addition, the incoherent waves
generated by inelastic scattering will individually also be scattered back and
forth and generate coherent sets of their own. The situation which
combines coherence from elastic scattering and incoherence and
probability from inelastic scattering is formally described by a
density matrix. The density matrices resulting from all the
incident plane waves which are incoherent to one another should be
added to obtain the net density matrix.

In this paper, we do not undertake the calculation of the density
matrix of the sample. Rather, we assume that the density matrix
has been calculated. We did include the macroscopic contacts in
the problem, since they are needed in determining the density
matrix of the sample in the Non Equilibrium Green Function-Landauer approach\cite{datta}. The objective
of this paper is to calculate the total tunneling current based on
the density matrix of the sample and properties of the STM tip.
Once the total tunneling current is obtained with the voltage
applied on the tip as a parameter, by setting that current to zero
one implicitly determines STP measurement result.

\subsection{Total tunneling current}
In calculating the total tunneling current, we have the density
matrix of the sample, denoted by $\hat{\varrho}$, and describe the
STM tip with a Fermi-Dirac distribution function:
\[
f(E_{c,\vec p}-eV_t-\mu)=\frac{1}{e^{\beta(E_{c,\vec p}-eV_t-\mu)}+1}\ ,
\]
with $\{|\chi_{\vec p}\rangle\}$ as its basis set. Here $V_t$ is
the voltage applied on the STM tip, $E_{c,\vec p}$ is the chemical
eigen energy of state $|\chi_{\vec p}\rangle$, and $\mu$ is the
electrochemical potential of the tip.

To derive the expression for the total tunneling current, we start
by diagonalizing the density matrix $\hat{\varrho}$ of the sample
. After diagonalization, the density matrix reads:
\begin{equation}\label{total1}
\hat{\bm{\rho}}_\psi=\left(\begin{matrix}\ddots&&\cdots& &
\cdots\\&\rho_{\psi\vec k\vec k}& &0
\\\vdots& & \ddots& &\vdots\\&0&&\rho_{\psi\vec k'\vec
k'}&&\\\cdots&&\cdots&&\ddots
\end{matrix}\right)\ ,
\end{equation}
where $\rho_{\psi\vec k\vec k}=\langle\psi_{\vec
k}|\hat{\varrho}|\psi_{\vec k}\rangle$. Here we use
$\hat{\varrho}$ for the abstract density matrix, and use
$\hat{\bm{\rho}}_\psi$ for the matrix form of the density matrix
under the basis set $\{|\psi_{\vec k}\rangle\}$. In this
particular case, this basis set diagonalizes the density matrix.

The basis set $\{|\psi_{\vec k}\rangle\}$ consists of wave
functions that are incoherent to each other in the problem, and
the diagonal elements in the density matrix $\hat{\bm{\rho}}_\psi$
have physical meanings that the probability of finding an electron
in state $|\psi_{\vec k}\rangle$ is $\rho_{\psi\vec k\vec k}$.
Note that elastic scattering only scatters states with a given
eigen electrochemical energy into states with the same eigen
electrochemical energy, thus one can choose $\{|\psi_{\vec
k}\rangle\}$ such that they are all energy eigen states, with
$E_{ec,\vec k}$ as their eigen electrochemical energies. Since we
are using chemical energies for states in the tip and
electrochemical energies for states in the sample, the subscripts
``s'' and ``t'' in the denotation of energy levels for sample and tip
have been omitted.

\begin{figure}
\begin{center}
\includegraphics[width=2in,bb=10 20 150 150]{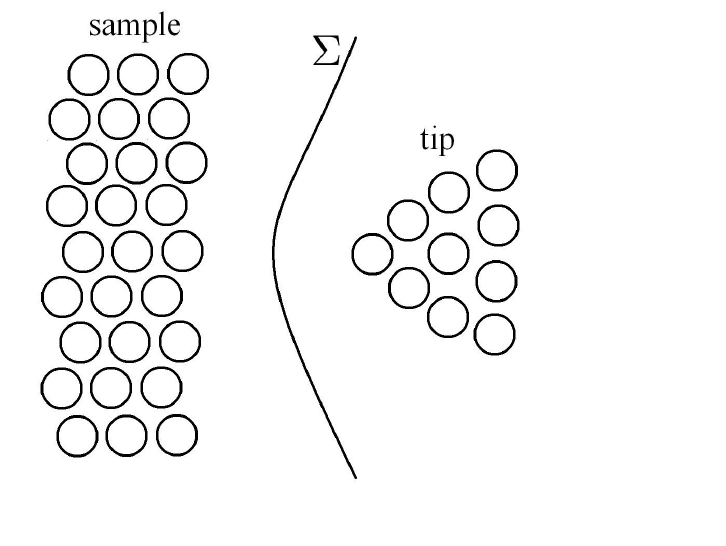}
\caption{Calculation of the tunneling current. The tunneling
matrix element between one state of the sample and one state of
the tip is evaluated by a surface integration over the surface
$\Sigma$ between the sample and the tip (see equation (\ref{total2})). Refer to reference
\onlinecite{stmbook} for details. This figure is redrawn in complete analogy to Figure 3.1 in this reference. Here the circles represent atoms of the sample (left) or the tip (right), and the surface $\Sigma$ is an imaginary surface in vacuum over which the integration in equation (\ref{total2}) is performed.}\label{tunnelmatrix}
\end{center}
\end{figure}

As shown in Figure \ref{tunnelmatrix}, the tunneling current
between one state $|\psi_{\vec k}\rangle$ of the sample and one
state $|\chi_{\vec p}\rangle$ of the tip is calculated through
evaluating a surface integral over $\Sigma$ between the sample and
the tip, and reads\cite{stmbook}:
\[
I_{\vec k,\vec p}=\frac{4\pi e}{\hbar}\big[\rho_{\psi\vec k\vec
k}-f(E_{c,\vec p}-eV_t-\mu)\big]\delta(E_{ec,\vec k},E_{c,\vec
p}-eV_t)
\]
\begin{equation}\label{total2}
\qquad\times\left(\frac{\hbar^2}{2m}\right)^2\left|\int_{\Sigma}(\psi_{\vec
k}\vec\nabla \chi^*_{\vec p}-\chi^*_{\vec p}\nabla \psi_{\vec
k})\cdot\mbox{d}\vec S\right|^2\ .
\end{equation}
The tunneling current resulting from different channels are
additive, because both the states of the sample and the states of
the tip are incoherent with one another. Thus the total tunneling
current is:
\[
I=\sum_{\vec k,\vec p}\Bigg\{\frac{4\pi
e}{\hbar}\big[\rho_{\psi\vec k\vec k}-f(E_{c,\vec
p}-eV_t-\mu)\big]\delta(E_{ec,\vec k},E_{c,\vec p}-eV_t)
\]
\begin{equation}\label{total3}
\qquad\times\left(\frac{\hbar^2}{2m}\right)^2\left|\int_{\Sigma}(\psi_{\vec
k}\vec\nabla \chi^*_{\vec p}-\chi^*_{\vec p}\nabla \psi_{\vec
k})\cdot\mbox{d}\vec S\right|^2\Bigg\}
\end{equation}

The tunneling matrix element:
\begin{equation}\label{total4}
M_{\vec k\vec p}\equiv\delta(E_{ec,\vec k},E_{c,\vec
p}-eV_t)\frac{\hbar^2}{2m}\int_{\Sigma}(\psi_{\vec k}\vec\nabla
\chi^*_{\vec p}-\chi^*_{\vec p}\nabla \psi_{\vec
k})\cdot\mbox{d}\vec S
\end{equation}
can be written in another way if one expands the wave function of
the tip into spherical harmonics (see chapter 3.2 of reference
\onlinecite{stmbook}):
\begin{equation}\label{total5}
\chi_{\vec p}(\vec r)=\sum_{l,m}C_{lm,\vec
p}k_l(\kappa\rho)Y_{lm}(\theta,\phi)\ ,
\end{equation}
where besides the familiar denotation of $Y_{lm}$ as the spherical
harmonic functions, $k_l$ is the $l^{\text{\scriptsize th}}$
spherical modified Bessel function of the second kind. There
exists a general derivative rule (chapter 3.4 of reference
\onlinecite{stmbook}) that the density matrix element (\ref{total4}) can
be written as a linear operation of the sample state only,
evaluated at the center of curvature $\vec r_0$ of the tip which
is also the origin for expanding the tip wave functions:
\begin{equation}\label{total6}
M_{\vec k\vec p}=\left(\mathcal F_{(\vec p,V_t)}\psi_{\vec
k}\right)|_{\vec r_0}\ .
\end{equation}
For instance, if the tip wave function is well approximated by a
$s$-wave, then:
\[
\mathcal F_{(\vec p, V_t)}=\frac{2\pi C_{\vec p}\hbar^2}{\kappa
m}\delta(E_{ec,\vec k},E_{c,\vec p}-eV_t)\ ,
\]
which is a pure number. As another example, for a $p_z$-wave tip
state, one has:
\[
\mathcal F_{(\vec p, V_t)}=\frac{2\pi C_{\vec p}\hbar^2}{\kappa
m}\delta(E_{ec,\vec k},E_{c,\vec p}-eV_t)\frac{\partial}{\partial
z}\ .
\]
When the pure state $\chi_{\vec p}$ is a combination of the s, p,
d $\cdots$ waves, the corresponding operators are additive.

%\newpage

The total tunneling current can be expressed:
\begin{equation}\label{total7}
I=\sum_{\vec k,\vec p}\frac{4\pi e}{\hbar}\big[\rho_{\psi\vec
k\vec k}-f(E_{c,\vec p}-eV_t-\mu)\big]\bigg|\left(\mathcal
F_{(\vec p, V_t)}\psi_{\vec k}\right)|_{\vec r_0}\bigg|^2\ ,
\end{equation}
where the Kronecker delta symbol has been absorbed in the operator
$\mathcal F_{(\vec p,V_t)}$, and these operators are well defined
only between energy eigen states. From the point of view of the
heuristic approach in Section \ref{stp}, $\rho_{\psi\vec k\vec k}$
serves as the distribution function, and $\bigg|\left(\mathcal
F_{(\vec p, V_t)}\psi_{\vec k}\right)|_{\vec r_0}\bigg|^2$ serves
as the magnitude of tunneling matrix element, and the averaging of
the tunneling matrix element is effectively defined such that
equation (\ref{heuristic2}) equals equation (\ref{total7}). Note that the densities of states of both the sample and the tip have been reflected in the summation over all states of the tip and all states of the sample.

The above expression for the total tunneling current requires
diagonalization of the density matrix of the sample. To obtain an
abstract expression, one needs to rewrite it into a form which is
a product of a matrix with the density matrix of the sample. When
only states that are electrochemical energy eigen states are used
in the basis sets, say $\{|\phi_{\vec k}\rangle \}$, with density
matrix of the sample in the form:
\begin{equation}\label{total8}
\hat{\bm{\rho}}_\phi=\left(\begin{matrix}\ddots&&\cdots& &
\cdots\\&\rho_{\phi\vec k\vec k}& &\rho_{\phi\vec k\vec k'}
\\\vdots& & \ddots& &\vdots\\&\rho_{\phi\vec k'\vec k}&&\rho_{\phi\vec k'\vec k'}
&&\\\cdots&&\cdots&&\ddots
\end{matrix}\right)\ ,
\end{equation}
this matrix-product form of the expression is:
\begin{widetext}
\begin{equation}\label{total9}
I=\sum_{\vec
p}\frac{4\pi e}{\hbar}tr\bigg[\vec \phi_0^{\dag}\vec
\phi_0[\hat{\bm{\rho}}_{\phi}-f(E_{c,\vec p}-eV_t-\mu)]\bigg]=\frac{4\pi e}{\hbar}tr\left[\big(\sum_{\vec p}\vec
\phi_0^{\dag}\vec
\phi_0\big)\hat{\bm{\rho}}_{\phi}-\left(\sum_{\vec p}f(E_{c,\vec
p}-eV_t-\mu)\vec \phi_0^{\dag}\vec \phi_0\right)\right]\ ,
\end{equation}
\end{widetext}
where:
%\newpage

\begin{widetext}
\begin{equation}
\vec \phi_0=\left(\begin{matrix}\mathcal F_{(\vec
p,V_t)}\phi_{\vec k1}(\vec r_0),&\cdots,&\mathcal F_{(\vec
p,V_t)}\phi_{\vec ki}(\vec r_0),&\cdots,&\mathcal F_{(\vec
p,V_t)}\phi_{\vec kn}(\vec r_0)\end{matrix}\right)\ .
\end{equation}
\end{widetext}
%and:
%\[
%\vec \phi_0^{\dag}=\left(\begin{matrix}\mathcal F_{(\vec
%p,V_t)}\phi^*_{\vec k1}(\vec r_0)\\\vdots\\\mathcal F_{(\vec
%p,V_t)}\phi^*_{\vec ki}(\vec r_0)\\\vdots\\\mathcal F_{(\vec
%p,V_t)}\phi^*_{\vec kn}(\vec r_0)\end{matrix}\right)\ .
%\]
Here, the order of stacking $\phi_{\vec k}$'s for the symbols
$\vec \phi_0^{\dag}$ and $\vec \phi_0$ is the same as that was
adopted in writing the density matrix $\hat{\bm{\rho}}_{\phi}$.
Note that the wave functions in the basis set $\{|\phi_{\vec
k}\rangle\}$ are all electrochemical energy eigen functions, which
makes $\mathcal F_{(\vec p,V_t)}$ well defined.

Although the above results are explicitly written down with the
basis set $\{|\phi_{\vec k}\rangle\}$ which includes only energy
eigen functions, one does not need this eigen function
requirement. The matrices $\sum_{\vec p}\vec \phi_0^{\dag}\vec
\phi_0$ and $\sum_{\vec p}f(E_{c,\vec p}-eV_t-\mu)\vec
\phi_0^{\dag}\vec \phi_0$ have the same transformation rules as
other operators (and density matrices) in the Hilbert space. Under
other basis sets, in general one does not have an explicit expression that
links the matrix elements to the value of the sample wave function
at a certain point (except for the special case discussed shortly in next subsection).

In equation (\ref{total9}), the tip and sample properties are entangled by the $\delta$-symbols, and we were not able to find an explicit expression for the measured STP potential $\mu$ although it is implicitly defined by setting equation (\ref{total9}) to be zero. We now introduce some further simplifications that make explicit expressions possible.
\subsection{Explicit expressions for measured potential}\label{explicit}

In this subsection we assume that the tip is featureless, in the sense that: a) the tip has a uniform density of states $N_t$ over the energy range that we are interested in; and b) the operator defined in equation (\ref{total6}) equals:
\begin{equation}\label{explicit1}
\mathcal F_{(\vec p, V_t)}=\delta(E_{ec,\vec k},E_{c,\vec p}-eV_t)\mathcal F_0\ ,
\end{equation}
i.e., the $\vec p$ dependence is only in the $\delta$-symbol.

Under the above simplifications, in equation (\ref{total9}) the sum over $\vec p$ can be explicitly carried out and one obtains:
\begin{equation}\label{explicit2}
\sum_{\vec p}\vec \phi_0^{\dag}\vec \phi_0=N_t\vec \phi_{00}^{\dag}\vec \phi_{00}\ ,
\end{equation}
\begin{equation}\label{explicit3}
\sum_{\vec
p}f(E_{c,\vec p}-eV_t-\mu)\vec \phi_0^{\dag}\vec \phi_0=N_t\vec \phi_{00}^{\dag}\vec \phi_{00}\hat{\bm\rho}_{\phi,FD}(\mu)\ ,
\end{equation}
where:
\begin{widetext}
\begin{equation}\label{explicit9}
\vec \phi_{00}=\left(\begin{matrix}\mathcal F_0\phi_{\vec k1}(\vec r_0),&\cdots,&\mathcal F_0\phi_{\vec ki}(\vec r_0),&\cdots,&\mathcal F_0\phi_{\vec kn}(\vec r_0)\end{matrix}\right)\ ,
\end{equation}
\end{widetext}
\begin{equation}\label{explicit4}
\hat\varrho_{FD}(\mu)=\frac{1}{1+e^{\beta(\hat H_s-\mu)}}\ ,
\end{equation}
and $\hat H_s$ is the Hamiltonian of the sample.

Note that the above equations are basis-transformation-independent, hence we do \emph{not} require the basis wave functions to be energy eigen functions for the expressions in this subsection. The matrix form of the total tunneling current is then:
\begin{equation}\label{explicit5}
I=\frac{4\pi e}{\hbar}N_t\cdot tr\left[\vec
\phi_{00}^{\dag}\vec
\phi_{00}\left(\hat{\bm{\rho}}_{\phi}-\hat{\bm{\rho}}_{\phi,FD}(\mu)\right)\right]\ ,
\end{equation}

%MAC RESTARTED HERE.
%
%GOING BACK A STEP, WOULD IT BE POSSIBLE TO ADD A DISCUSSION AFTER THE EQUATION ABOVE THAT RELATES ITS FORM TO THAT FOUND IN THE HEURISTIC TREATMENT.  IT LOOKS PRETTY SIMILAR, ONLY MORE PRECISE.

It is noteworthy to compare the above equation to equation (\ref{heuristic2}). One can immediately see the resemblance of the two equations, except that the density of states of the sample is implicitly included in the summation of the trace operation. One can also see that it is the difference of the density matrix that STP measures (with a weight), thus in order to express things in terms of effective distribution functions, a mean matrix element value should be used and that value is expected to change over space (as seen in the definition of $\vec \phi_{00}$ in equation (\ref{explicit9})).

\paragraph{Subcase I}
In this subcase, we consider a situation where temperature is high compared to deviations from thermal equilibrium due to the applied current. For example, in Figure \ref{disfunc} for the heuristic understanding of STP, we need to have $k_BT>>el_{in}E_0$, rather than a sharp transition in distribution function implying $T\approx0$. We assume the density matrix is very close to the thermal equilibrium with electrochemical potential $\mu_0$, and measured STP potential is $\mu_0+\Delta\mu$, then:
\[tr\left[\vec
\phi_{00}^{\dag}\vec
\phi_{00}\left(\hat{\bm{\rho}}_{\phi}-\hat{\bm{\rho}}_{\phi,FD}(\mu_0)\right)\right]
\]
\begin{equation}\label{explicit6}
=tr\left[\vec
\phi_{00}^{\dag}\vec
\phi_{00}\left(\hat{\bm{\rho}}_{\phi,FD}(\mu_0+\Delta\mu)-\hat{\bm{\rho}}_{\phi,FD}(\mu_0)\right)\right]
\end{equation}
Noticing:
\[
\frac{1}{1+e^{\beta(E-\mu_0-\Delta\mu)}}-\frac{1}{1+e^{\beta(E-\mu_0)}}
\]
\[
\approx\frac{\beta}{2+e^{-\beta(E-\mu_0)}+e^{\beta(E-\mu_0)}}\Delta\mu\ ,
\]
we get:
\begin{equation}\label{explicit7}
\Delta\mu=\frac1\beta\frac{tr\left[\vec
\phi_{00}^{\dag}\vec
\phi_{00}\left(\hat{\bm{\rho}}_{\phi}-\hat{\bm{\rho}}_{\phi,FD}(\mu_0)\right)\right]}{tr\left[\vec
\phi_{00}^{\dag}\vec
\phi_{00}\hat{\bm{\rho'}}_{\phi,FD}(\mu_0)\right]}\ ,
\end{equation}
where:
\[
\hat{\varrho'}_{FD}(\mu_0)=\frac{1}{2+e^{-\beta(H_s-\mu_0)}+e^{\beta(H_s-\mu_0)}}\ .
\]

%IT OCCURS TO ME IN THIS SECTION AND THE EQUATION FOR DELTA MU ABOVE IN PARTICULAR THAT WE SHOULD AT LEAST DISCUSS THE POTENTIAL FOR INTERPRETING EXPERIMENT USING THESE RELATIONSHIPS.  HERE AND BELOW YOU COMMENT ON HOW HARD IT IS TO CALCUATE THE DENSITY MATRIX TO COMPARE WITH EXPERIMENT, BUT WHAT CAN BE SAID WITHOUT SUCH CALCUATIONS, OR TO GUIDE SUCH CALCULATIONS, OR TO EXTRACT PHYSICS PURELY FROM EXPERIMENT.

\paragraph{Subcase II}
In this subcase, we assume that both thermal smoothing and nonequilibrium modification of the distribution function corresponds to a small reciprocal vector change $\Delta\vec k$ compared to $\vec k_F$ and all other characteristic reciprocal vectors. A somewhat cruder approximation reads:
\[
\frac{1}{1+e^{\beta(E-\mu_0-\Delta\mu)}}-\frac{1}{1+e^{\beta(E-\mu_0)}}
\]
\[
\approx\delta(E-\mu_0)\Delta\mu\ ,
\]
hence one obtains:
\begin{equation}\label{explicit8}
\Delta\mu=\frac{tr\left[\vec
\phi_{00}^{\dag}\vec
\phi_{00}\left(\hat{\bm{\rho}}_{\phi}-\hat{\bm{\rho}}_{\phi,FD}(\mu_0)\right)\right]}{tr\left[\vec
\phi_{00}^{\dag}\vec
\phi_{00}\hat{\bm{\rho''}}_{\phi,FD}(\mu_0)\right]}\ ,
\end{equation}
where:
\[
\hat{\varrho''}_{FD}(\mu_0)=\delta(H_s-\mu_0)\ .
\]

The results above constituted a general theory of STP that is valid on all length scales and puts coherent and incoherent processes on an equal footing through the use of the density matrix.  The final results from which the measured potential is implicitly determined are equations (\ref{total3}) or (\ref{total9}), which are equivalent.   The result is formal, however, and not as useful as would be desirable because it requires knowledge of the density matrix over the entire sample -- a formidable problem indeed.  As we argue on physical grounds in the next section, some form of local density matrix would be desirable and seems  possible to us, although we do not claim to provide a rigorous proof.  Put more strongly, a formulation in terms of local properties will be necessary in order to make more quantitative interpretation of STP result practical.

%READ THE FOLLOWING SECTIONS CAREFULLY.  I HAVE MADE SOME CHANGES IN THE TEXT TO TIGHTEN THE NARRATIVE AND IN THE CASE OF THE LOCAL DENSITY MATRIX TO USE THE CONDITIONAL TENSE, AS SEEM APPROPRIATE TO ME GIVEN OUR (AT LEAST) MY THEORETICAL LIMITIATIONS.

\section{Local density matrix}\label{local}

As argued above, the simple model posed here requires
including both contacts providing the current and large areas of
the sample that are far away from the region where STP
measurement is performed, this is clearly unphysical.  Moreover, the density matrix of the
sample is very difficult to calculate numerically due to the large  number of
degrees of freedom.  Equally important, ideally one would like a theory with which one can uncover relevant physics of transport without a priori knowledge of the density matrix of the sample.
A proper theory in this case should  only involve wave functions in
the vicinity of the probed area, and involve defects and
material characteristics in roughly the same area. The measurement
result in this case should be determined by these parameters and a
proper boundary condition, for example the mean current flowing
near the region of the sample. We argue here that to deal with these issues it is better to describe the
situation with local density matrices defined for each point on
the sample, with the entries of the local density matrices
consisting of only local wave functions.

\subsection{Outline of local density matrix}

Let us outline here how such a theory might be constructed.  To begin, one needs electron wave functions that
only extend locally. The energy eigen states that we used above to
calculate the total tunneling current are extended throughout the
sample, at least in the case where there are no defects in the
sample. Using this basis set, all the states involved need to be
counted in expression (\ref{total9}), because the wave functions
have non-zero magnitude at the center of curvature of the tip
$r_0$. In order to isolate electrons far away from the measurement
area, we suggest to go to other basis sets where only local
wave functions are involved. In principle the density matrix of the sample is the same as when we use energy eigen states as basis states, however, a new local basis set will lead to an approximation that reduces the dimension of the density matrix, as we see below.

\begin{figure}
\begin{center}
\includegraphics[width=2.5in,bb=15 0 100 80]{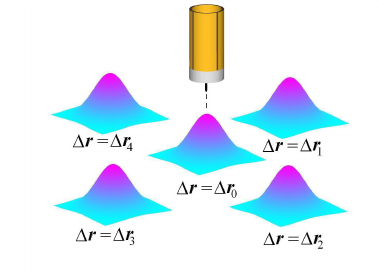}
\caption{Schematic of the new basis set that we propose to use in
order to construct local density matrix of the sample. In this
figure, only the $\Delta\vec r$ degree of freedom is shown. The
new basis set is defined with reference to the position of the STM
tip $\vec r$, with the states centered around $\vec r+\Delta\vec
r_i$. The states have finite wave function values at position
$\vec r$ when $\Delta\vec r_i$ is small, and states far away with
$\Delta\vec r_i$ large have negligible wave function values. When
the position of the STM tip changes, the set $\{\Delta\vec r_i\}$
remains the same, hence the central locations $\{\vec r+\Delta\vec
r_i\}$ changes, and the basis set changes.}\label{newset}
\end{center}
\end{figure}

We propose the use of a different basis set for each point on the
sample, see also Figure \ref{newset}. The basis set $\{|\xi_{\vec
r+\Delta\vec r_i,\vec k}\rangle\}$ consists of electron wave
packets that are centered around different positions $\vec
r+\Delta\vec r_i$ with wave vector centered around $\vec k$. Here,
the set $\{\Delta\vec r_i\}$ and $\{\vec k\}$ and the wave
function forms are the same for each point, but the vector $\vec
r$, which is the position where the STM tip probes, changes at
different measurement positions. The matrices that are transformed from
$\sum_{\vec p}\vec \phi_0^{\dag}\vec \phi_0$ and $\sum_{\vec
p}f(E_{c,\vec p}-eV_t-\mu)\vec \phi_0^{\dag}\vec \phi_0$ are thus
the same for different positions, except for the lateral change in
$\vec r$. While being incorrect in a strict quantum mechanical context, with this new basis set, one can intuitively talk about electrons at certain positions with certain wave numbers.

Intuitively, the new basis set needs to have $\Delta\vec r_i$
covering the full range of the sample, in order to be a complete
basis set. The matrix form of the density matrix of the sample
$\langle\xi_{\vec r+\Delta\vec r_i,\vec k}|\hat\varrho|\xi_{\vec
r+\Delta\vec r_j, \vec k'}\rangle$ changes at different positions
because the basis set changes.

The simplification from this approach is that density matrix elements
involving wave packets far away ($\Delta\vec r_i$ sufficiently
large) should not affect measurement result at position $\vec r$.
Note that the density matrix element for these states will be
finite, but because the region of finite amplitude for the wave
function is far from $\vec r$, one could expect that both matrices
$\sum_{\vec p}\vec \phi_0^{\dag}\vec \phi_0$ and $\sum_{\vec
p}f(E_{c,\vec p}-eV_t-\mu)\vec \phi_0^{\dag}\vec \phi_0$ in
equation (\ref{total9}) have zero elements involving these
states\footnote{This is obtained from physical intuition, which
needs to be verified once the basis set is chosen. In fact, the
intuition that zero wave function value implies zero tunneling
current comes from energy eigen states, but the new basis set does
not consist of energy eigen states in the strict sense. In the special case discussed in subsection \ref{explicit} with a featureless tip, this physical intuition is formally proven.}. With
this simplification in mind, one might make an approximation and discard all states with
sufficiently large $\Delta\vec r_i$, and reduce the problem to one
that only involves electron wave packets near to the STM tip
position $\vec r$. This reduced matrix form that changes with
position is what we define as the local density matrix.

One could thus interpret that what STP probes in the relevant non-equilibrium conditions is the local density matrix at certain positions on the sample.

We note that if the total tunneling current was to only relate to
the local density matrix at one point, the STM tip needs to be
sharp, with its radius of curvature smaller than the lengthscale
of variation of the local density matrix. Only in that case could
one choose a surface $\Sigma$ in equation (\ref{total2}) that is
localized in space.

The concept of using basis sets with localized wavefunctions is not new, for example, codes based on linear combination of atomic orbitals (LCAO) are common in electron transport calculations\cite{LCAO1,LCAO2}. While using these basis sets do naturally lead to local density matrices in our notion, we would like to point out that these existing usages are mostly on small, or even molecular, samples; the calculational strategy for a proper calculation on large samples, as we will soon discuss, has not been worked out. Hence by introducing the concept of local density matrix we not only provide a theoretical convenience, but also pose a new problem in mesoscopic transport.

\subsection{Possible further developments}
The local density matrix is defined on a basis set
$\{|\xi_{\vec r+\Delta\vec r_i,\vec k}\rangle\}$ which consists of
only wave packets near to the position of STM tip $\vec r$. The
matrices that are transformed from $\sum_{\vec p}\vec
\phi_0^{\dag}\vec \phi_0$ and $\sum_{\vec p}f(E_{c,\vec
p}-eV_t-\mu)\vec \phi_0^{\dag}\vec \phi_0$ in equation
(\ref{total9}) are both determined by transformation rules in the
Hilbert space, and, once computed, are fixed except for the
position $\vec r$ for different positions that the STM tip probes. In the special case described in subsection \ref{explicit}, the two matrices above are given explicitly by equation (\ref{explicit9}), no matter what basis set one uses.

We propose that the approach to calculate the local density matrix be different than the path of its definition. Since the local density matrix is spatially variant, there should be a
controlling differential equation that determines its variation
over space. One can thus pose a problem with proper boundary
conditions that reflects the mean current around the probed area,
and need not concern the current-providing contacts or the sample
area far away from the probed area. In other words, although the derivation of the controlling equations is based on the density matrix of the sample, in the actual calculation only the local density matrix is relevant, and one does not even need to know the density matrix of the sample. The detailed form of the differential equations and boundary conditions should depend on what basis set one chooses and what further approximations one makes. We do not address these computational challenges in this paper.  They require proper theorists.  On the other hand, below we provide preliminary thoughts of possible further simplification procedures.

The choice of the new basis set is clearly critical in the
problem. The size of the wave packets in space should be small
compared to macroscopic lengths in order to make the
simplification with local density matrix feasible; meanwhile this
size should be large compared to $\lambda_F$, because of
uncertainty principle and the need to keep the energy uncertainty
small. One can thus imagine two limits to simplify the problem:
one is to make relatively large wave packets, thus including more
states in the local density matrix, and making the uncertainty
in energy small; the other is to make relatively small wave
packets, which allows to include few states in the local density
matrix, but at the expense of large energy uncertainty for each
basis state.

The matrices that are transformed from $\sum_{\vec p}\vec
\phi_0^{\dag}\vec \phi_0$ and $\sum_{\vec p}f(E_{c,\vec
p}-eV_t-\mu)\vec \phi_0^{\dag}\vec \phi_0$ in equation
(\ref{total9}) depend on how one chooses the new basis set at each
point. One can calculate these matrices according to the
transformation rules, or in the case where large wave packets are
adopted, one may assume that each wave packet is an approximate
energy eigen state and apply these two matrices in the explicit
form.

To simplify the problem without losing important information, it
is reasonable to assume that all the operators $\mathcal F_{(\vec
p,V_t)}$ are the same for the states in the tip as in the special case of subsection \ref{explicit}, viewing the tip
as part of the instrument and being controlled by the
experimentalist. For example, since only nanometer scale signals
are being probed, one can assume that the tip is an $s$-wave tip
with the same constant in the matrix element calculation. This
approximation was adopted in Chu \& Sorbello's
paper\cite{sorbello}. Under this approximation the local density matrix approach is also better solidified.

\section{Limiting cases}\label{limiting}
\subsection{Sample in equilibrium}
When the sample does not bear any current, i.e., is in
equilibrium, after diagonalization the density matrix of the
sample will become a Fermi-Dirac distribution, i.e., suppose the
basis set $\{|\psi_{\vec k}\rangle\}$ diagonalizes $\hat\varrho$,
one obtains from equation (\ref{total3}):
\begin{widetext}
\begin{equation}\label{limit1}
I=\sum_{\vec k,\vec p}\Bigg\{\frac{4\pi e}{\hbar}\big[f(E_{ec,\vec
k}-\mu_s)-f(E_{c,\vec p}-eV_t-\mu_t)\big]\delta(E_{ec,\vec
k},E_{c,\vec p}-eV_t)\bigg|\left(\mathcal F_{(\vec p, V_t)}\psi_{\vec
k}\right)|_{\vec r_0}\bigg|^2\Bigg\}\ .
\end{equation}
\end{widetext}
Note that the Kronecker delta symbol requires the electrochemical
energies for the sample state and the tip state to be the same.
Under this condition and when the temperatures of the tip and the
sample are the same, if $\mu_s>\mu_t$, all the terms in the above
equation will be positive, hence the total tunneling current
becomes positive; similarly if $\mu_s<\mu_t$, the total tunneling
current becomes negative. The only possibility for the total
tunneling current to be zero is if $\mu_s=\mu_t$, regardless where
the tip is. This is expected from physical consideration. When the
temperatures of the tip and the sample are different, the above
conclusion is no longer true, but this is expected as a thermal
electric effect.
\subsection{Homogeneous sample with no defect}
Suppose the sample does not have any defects, namely, there is only inelastic scattering and no
elastic scattering in the sample. Also suppose that the current
is uniform on the sample. In a jellium model, from translational symmetry of the sample
(neglecting effects due to the macroscopic contacts), one expects
that the local density matrix should be the same throughout the
probed area, except for the effect from a linear electrostatic
potential distribution reflecting the fact that there is a
constant current and there exists inelastic scattering. In this
case, the local density matrix is described by a distribution function (no off-diagonal elements),
because electrons at different eigen states are incoherent to each other.
In the linear response limit, the distribution function reads:
\[
f(\vec k,\vec r)\propto\frac1{e^{\beta(E_{\vec
k}-\mu_{\hat\theta,\vec r})}+1}
\]
\begin{equation}\label{limit2}
\mu_{\hat\theta,\vec r}=e\vec E\cdot\vec r+\mu-el_{in}\vec
E\cdot\hat\theta\ .
\end{equation}
For each direction of momentum there exist an effective chemical
potential $\mu_{\hat\theta}$. This distribution function leads to the distribution function with electrochemical energy as a variant in equation (\ref{heuristic1}). In this case, in STP measurement the difference of data at different points comes from the $-\vec
E\cdot\vec r$ term.
\subsection{Landauer resistive dipole}
Suppose there is only one defect on the otherwise homogeneous
sample, or assume the defects are far away from each other
compared to $l_{in}$. Far away from the defect(s), electrons are not coherent to one another due to inelastic scattering events after they are reflected by the defect(s). At these positions, the local density matrix
will again be described by a distribution function. In the linear response limit, the
distribution function will be very similar to equation
(\ref{limit2}) above, namely:
\[
f(\vec k,\vec r)\propto\frac1{e^{\beta(E_{\vec
k}-\mu_{\hat\theta,\vec r})}+1}
\]
\begin{equation}\label{limit3}
\mu_{\vec\theta,\vec r}=-e\phi(\vec
r)+\mu+el_{in}\nabla\phi\cdot\hat\theta\ ,
\end{equation}
where $\phi(\vec r)$ is the Landauer resistive dipole potential\cite{landauer}.
This demonstrates that STP can measure the Landauer resistive
dipole potential, which is expected from physical
consideration. Resistive dipoles have been observed by
experiments\cite{STP}. Equations (\ref{limit2}) and (\ref{limit3}) are similar to each other, because at the regions where Landauer derived the resistive dipole, the only practical difference is that the electrostatic potential is different than the previous limiting case due to the existence of the scattering center. Here $\phi(\vec r)$ is the dipole field plus the homogeneous electric field.
\subsection{Chu and Sorbello model}
Chu and Sorbello\cite{sorbello} calculated the STP measurement result much nearer to a defect than $l_{in}$. In the paper it is assumed that the electrons have a specific distribution function in $\vec k$ space before encountering the scattering center, and after being deflected by the scatterer, the electrons are not deflected again by the same scatterer. Interpreting their calculation in our context, they view most part of the sample as the current-providing electrodes that inject the incoming electrons with the specific distribution into the ``sample", and the sample size is effectively much smaller than $l_{in}$, centering around the defect. Once the electrons enter the sample they remain coherent
 in the presence of elastic scattering. The reservoir case, the background scatterers case and the
semiclassical barrier case are all describing the
distribution of the injected electrons. In our context, the density matrix of the sample, which is only a very small region near the defect, is
completely known:
\begin{equation}\label{limit4}
\varrho=\sum_{\vec k}f(\vec k)\left|\psi^{(+)}_{\vec
k}\right\rangle\left\langle\psi^{(+)}_{\vec
k}\right|\ ,%\propto\sum_{\vec k}\frac1{e^{\beta(E_{\vec
%k}-\mu_{\hat\theta})}+1}\left|\psi^{(+)}_{\vec
%k}\right\rangle\left\langle\psi^{(+)}_{\vec k}\right|\ ,
\end{equation}
here $\left|\psi^{(+)}_{\vec k}\right\rangle$ is the purely
coherent state resulting from one channel of injected electron
from the contacts.

If one considers the case where the part of the sample that is not
close to the defect is also measured by STP, the above density
matrix no longer describes the situation, that is to say, equation
(\ref{limit4}) does not describe all the electrons in the system.
However, it is conceivable that with a local density matrix
description, the local density matrix very near the defect does
look like equation (\ref{limit4}).

Using equation (\ref{explicit8}) and noticing that $f(E_{\vec k}-\mu_{\vec k})-f(E_{\vec k}-\mu)\approx\delta(E_{\vec k}-\mu)(\mu_{\vec k}-\mu)$, one obtains:
\begin{equation}\label{limit5}
\delta V_{STP}({\vec r}_0)=\frac{1/e\sum_{\vec k}|\psi_{\vec
k}^{(+)}({\vec r}_0)|^2\delta(E_{\vec k}-\mu)(\mu_{\vec
k}-\mu)}{\sum_{\vec k}|\psi_{\vec k}^{(+)}({\vec r}_0)|^2\delta(E_{\vec
k}-\mu)}\ ,
\end{equation}
which is equation (14) in Chu and Sorbello's paper.
%one recognizes $\mu_{\bm k}$ as the effective chemical potential,
%and $|\psi_{\bm k}^{(+)}(\bm{r}_0)|^2$ as the magnitude of
%tunneling matrix elements.
This illustrates the reason why
tunneling matrix elements should not be assumed to have the same
magnitude in writing down the total tunneling current, as the
variation of this magnitude is the interference term. It can also
be seen from this expression that the density matrix of the sample
determines how and where this variation occurs.

%It is also worthwhile to note that in Chu \& Sorbello's model, the
%electrochemical potential (for example defined as effective
%electrochemical potential averaged for all channels) varies with
%spatial coordinates only of order $(\kappa/\lambda_F)^2$ (where
%$\kappa$ is the screening length of the sample) compared to the
%variation in electrostatic potential as calculated in their
%equation (6), due to screening. However, STP measures a similar
%variation magnitude as the electrostatic potential due to changes
%in tunneling matrix elements. In other words, STP does {\it not}
%measure electrochemical potential in this case, rather, in a sense
%it measures electron wave functions.

%Gramespacher and
%B\"{u}ttiker\cite{buttiker1,buttiker2} viewed STP
%measurement as a subcase in their calculation. They discussed the
%case where there are more than two contacts, each one in
%equilibrium with a different electrochemical potential, and view
%the tunneling junction as a weakly coupled contact. Their
%consideration treats a situation similar to that of Chu \&
%Sorbello\cite{sorbello} in the sense that
%once an electron enters the sample, it remains completely coherent.
%All the incoherence effects are from the contacts.
\subsection{Atomic resolution in STP}
As we have emphasized, the tunneling matrix elements are dependent on electron wave function values, which change over space, leading to the quantum interference effects in the end result of STP. In normal STM mode, it is also the change of tunneling matrix elements within a unit cell that leads to atomic resolution of STM (and practically, in order that the atomic corrugations be measurable, the tip should usually be sharper than an s-wave tip, i.e., the operators $\mathcal F$ acting on the wave functions should not be as simple as a pure number, see also discussions in Chapter 7 of reference \onlinecite{stmbook}). It is then natural to ask the following question: does there exist atomic resolution (i.e., corrugations in the end result due to wave function value changes within a unit cell) in STP measurements?

Part of the answer to the previous question is simple. As we discussed in the first limiting case in this section, without a current (sample in equilibrium), there shall be atomic resolution in STM, but one will see no atomic corrugation in STP measurement because it is a constant throughout the sample. This demonstrates that if there is atomic resolution in STP, it should be different in origin than atomic resolution in STM mode. The answer to this question when there is a current on the sample becomes complicated. To demonstrate the nature of this problem, we investigate the STP measurement on a simple toy model, namely a one-dimensional tight-binding model with only one atomic orbital involved. Readers are referred to Chapter 10 of reference \onlinecite{AM} for the basic setup of tight-binding model.

We assume that the sample is one dimensional and is well described by a tight-binding model with only one atomic orbital, the wave function of which is $\phi(x)$. $\phi(x)$ is centered around $x=0$. The lattice constant of the sample is $a$. In the tight-binding model, the wave functions of the band is known:
\begin{equation}\label{tb1}
\psi_{k}(x)=\sum_ne^{ikna}\phi(x-na)\ ,
\end{equation}
where $k\in(-\frac\pi a,\frac\pi a)$, which is the first Brillouin zone. For our purpose, it suffices to assume that $\phi(x)$ is a s-level and to use nearest neighbor approximation, which lead to the band structure:
\begin{equation}\label{tb2}
E(k)=E_0-\gamma\cos(ka)\ ,
\end{equation}
where $E_0$ and $\gamma$ are two numbers related to the s-level energy and $\Delta U(x)$ which is the correction to the atomic Hamiltonian. To further simplify things, we assume that all operators $\mathcal F$ are the same for all states (i.e., approximation(\ref{explicit1}) stands, and the expression for total tunneling current is equation (\ref{explicit5})) for now.

\subsubsection{Sample with no defects}
\begin{figure}
\begin{center}
\includegraphics[width=2.2in,bb=0 0 613 750]{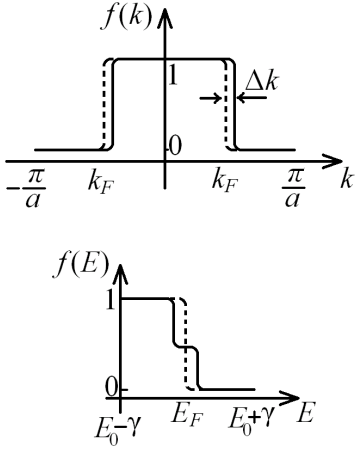}
\caption{Sketch of distribution function with respect to $k$ (upper panel) and $E$ (lower panel), both without current (dashed lines) and with a current (solid lines), in the tight-binding model without defects under consideration.}\label{tb}
\end{center}
\end{figure}

Suppose the sample has no defects, then the density matrix of the sample is diagonal under the basis of $\{\psi_k\}$. We can use a distribution function $f(k)$ to describe things, and the distribution function with respect to energy $f(E)$ can be calculated from $f(k)$. If we ignore the thermal broadening, the distribution function with respect to both $k$ and energy is sketched in Figure \ref{tb}.

The total tunneling current in this case reads:
\begin{equation}\label{tb3}
I=\frac{4\pi e}{\hbar}N_t\cdot \sum_{k}\left|\mathcal F_0\psi_k(x_0)\right|^2\left(f(k)-f_{FD}(k,\mu)\right)\ ,
\end{equation}
where $f_{FD}(k,\mu)$ is the Fermi-Dirac distribution function with chemical potential $\mu$ (dashed curve in lower panel of Figure \ref{tb} but has a freedom to move in the horizontal direction depending on the value of $\mu$). The question then becomes: does the value of $\mu$ that makes equation (\ref{tb3}) equal to zero have a spatial dependence (as $x_0$ changes within the unit cell)?
\paragraph{$x_0$ close to origin}
When we consider the STP measurement results in the vicinity of the atomic orbital, only the $n=0$ component in the definition of $\psi_k(x)$ in equation (\ref{tb1}) is relevant, thus the total tunneling current becomes:
\[
I\propto\sum_{k}\left|\mathcal F_0\phi(x_0)\right|^2\left(f(k)-f_{FD}(k,\mu)\right)
\]
\begin{equation}\label{tb4}
=\left|\mathcal F_0\phi(x_0)\right|^2\sum_{k}\left(f(k)-f_{FD}(k,\mu)\right)\ .
\end{equation}
Hence there is no spatial dependence in $\mu$. This is in contrast to atomicly resolved corrugations in STM mode, in which case the change in $\left|\mathcal F_0\phi(x_0)\right|^2$ indicates a spatial change in the measured sample height.
\paragraph{$x_0$ close to $a/2$}
Now consider the case where the probed point is not close to the vicinity of the atomic orbital, i.e., at least two (and in nearest neighbor approximation we consider only two, for example $n=0$ and 1) components in equation (\ref{tb1}) should be considered. The total tunneling current then reads:
\begin{widetext}
\[
I\propto\sum_{k}\left|\mathcal F_0\phi(x_0)+e^{ika}\mathcal F_0\phi(x_0-a)\right|^2\left(f(k)-f_{FD}(k,\mu)\right)
=\left(\left|\mathcal F_0\phi(x_0)\right|^2+\left|\mathcal F_0\phi(x_0-a)\right|^2\right)\sum_{k}\left(f(k)-f_{FD}(k,\mu)\right)
\]
\begin{equation}\label{tb5}
+\sum_{k}\left[\left(\mathcal F_0\phi(x_0)\right)^*\mathcal F_0\phi(x_0-a)e^{ika}+\mathcal F_0\phi(x_0)\left(\mathcal F_0\phi(x_0-a)\right)^*e^{-ika}\right]\left(f(k)-f_{FD}(k,\mu)\right)\ .
\end{equation}
\end{widetext}
It is reasonable to assume that $\left(\mathcal F_0\phi(x_0)\right)^*\mathcal F_0\phi(x_0-a)$ is real, then the second line of equation (\ref{tb5}) equals:
\begin{equation}\label{tb6}
2\left(\mathcal F_0\phi(x_0)\right)^*\mathcal F_0\phi(x_0-a)\sum_{k}\cos(ka)\left(f(k)-f_{FD}(k,\mu)\right)\ .
\end{equation}
When the current in the sample is small and the system is in linear response region, and if $\mu$ is the value that makes the first line of equation (\ref{tb5}) zero, then the range of $k$ for nonzero [$f(k)-f_{FD}(k,\mu)$] values consists of two very narrow regions centered around $-k_F$ and $k_F$, and because $\cos(k_Fa)=\cos(-k_Fa)$, expression (\ref{tb6}) equals zero, too. In other words, to this order there is still no spatial dependence in the measured STP potential.

If the current in the sample is large enough, such that $\Delta k$ in Figure \ref{tb} is significant, then the above argument breaks down and there will be a small corrugation in measured STP potential in the atomic resolution.

\subsubsection{Sample with defect(s)}
When the sample does have defect(s), under the basis $\{\psi_k\}$ the density matrix of the sample has off-diagonal elements. As we do not address the calculation of the density matrix, we only consider the case in which there is one off-diagonal element between states $k_0$ and $-k_0$ (degenerate states and $k_0$ is close to $k_F$). Suppose the off-diagonal element is $\delta$ in the $(k_0,-k_0)$ entry of the density matrix, and $\delta^*$ in the $(-k_0,k_0)$ entry. One only needs one additional term in equation (\ref{tb4}) or (\ref{tb5}) in this case:
\begin{equation}\label{tb7}
\Delta I=\mathcal F_0\psi_{k_0}(x_0)\left(\mathcal F_0\psi_{-k_0}(x_0)\right)^*\delta^*+\mbox{complex conjugate}
\end{equation}
With the same assumption above for the defect-free case that $\mathcal F_0\psi_{k_0}(x_0)\left(\mathcal F_0\psi_{-k_0}(x_0)\right)^*$ is real, we arrive at:
\begin{equation}\label{tb8}
\Delta I=2\mathcal F_0\psi_{k_0}(x_0)\left(\mathcal F_0\psi_{-k_0}(x_0)\right)^*Re(\delta)
\end{equation}

\paragraph{$x_0$ close to origin}
In this case, equation (\ref{tb4}) becomes:
\[
\sum_{k}\left(f(k)-f_{FD}(k,\mu)\right)+2Re(\delta)=0\ ,
\]
hence there is no spatial dependence in $\mu$.
\paragraph{$x_0$ close to $a/2$}
In this case, the correction to equation (\ref{tb5}) includes a term:
\[
2Re(\delta)\left(\mathcal F_0\phi(x_0)\right)^*\mathcal F_0\phi(x_0-a)\cos(k_0a)
\]
\[
\quad\times\left(f(k_0)-f_{FD}(k_0,\mu)\right)\ ,
\]
which does lead to a spatial dependence in $\mu$ which is proportional to $\delta$.

To sum up, when there are static defects in the sample, in the transition region between two adjacent atomic sites, there will be atomically resolved corrugations due to changes in wave function values, but this corrugation is a second order effect, as opposed to the first order effect in STM mode. If the sample does not have defects, hence no interference between states, the atomic corrugation is an even higher order effect.

Though the above results are obtained in a simplified toy model, we believe the conclusion bears some universality. It also demonstrates the nature of the STP potential problem, in that to first order the results are often trivial, in order to manifest effects from either wave function value changes or quantum interference, one often needs to go to a higher order. This is also part of the reasons that we are not able to provide estimates of the size of these effects without calculating the actual density matrices.

\section{Summary and discussion}
In this paper, we developed a viewpoint of STP measurement in the
framework of quantum transport. Equation (\ref{total9}) gives an
expression of the total tunneling current which is dependent on
the distribution function of the STM tip and the density matrix of
the sample. By setting this total tunneling current zero, one
implicitly determines the STP measurement result. The expression
is in a matrix product form such that a basis-set free definition
can be made. With some approximations, we also obtained explicit expressions for measured STP potential in the featureless tip case.

In order to get rid of the unphysical requirement of material
characteristics far away from the STP probing area in writing down
the density matrix of the sample, we proposed to use local density
matrices, defined in Section \ref{local}, as a varying parameter
over space, which is controlled by a certain differential equation
and a set of boundary conditions. This allows relating STP to
sample properties near the region where STP is performed.

In Section \ref{limiting} we provided limiting case calculations that demonstrate the usage of the theoretical description, mostly with given or assumed density matrices of the sample. In particular, we found that with the toy model of one dimensional tight-binding metal, atomic resolution in STP measurement is a high order effect. Following the derivation of this particular example, one can find that the null results are partly due to assumed symmetry in the Fermi surface (e.g., Equation (\ref{tb6}) being zero under the adopted assumptions) or in the operators $\mathcal F$ acting on the sample wave functions. We speculate that asymmetry in the Fermi surface with respect to $\vec k$, or in the tip wave function (hence operator $\mathcal F$) with respect to energy, could lead to lower order effects in STP measurement.

Comparing equation (\ref{total9}) with, for example, equation
(8.6.6) in reference \onlinecite{datta}, one can immediately see the
analogy. The matrices $\sum_{\vec p}\vec \phi_0^{\dag}\vec \phi_0$
and $\sum_{\vec p}f(E_{c,\vec p}-eV_t-\mu)\vec \phi_0^{\dag}\vec
\phi_0$ are in the same place where the scattering matrices are.
Although one needs to keep the subtle difference between
correlation function and density matrix in mind, these two
matrices can be viewed as a special kind of scattering matrix.

%We would like to emphasize here that the density matrix of the
%sample not only plays a similar role as a distribution function
%(especially after it is diagonalized), but also determines the
%tunneling matrix between the sample and the tip, or in other
%words, in the general notion of quantum transport, the scattering
%matrix between the sample and the tip . In the case of Chu and
%Sorbello's model$^{\text{\scriptsize\cite{sorbello}}}$, this
%convoluted nature of the entrance of density matrix is
%demonstrated by separating the effective distribution function and
%the tunneling matrix elements which are determined from the
%coherent wave functions. \textbf{be careful about this paragraph.
%the notion of tunneling matrix makes sense only when the density
%matrix is diagonalized. Otherwise, scattering matrices and the
%trace of the product of scattering matrices and density matrix is
%a more accurate description, with no explicit tunneling matrix
%elements in the expression. -- Weigang}

The STM tip as a special contact provides scattering which is
described by a scattering matrix related to the matrices mentioned
above. This implies that it will also act as an electron source,
the strength of which is dependent on the strength of scattering.
The electron source effect will change the density matrix of the
sample, or the local density matrix, as the tip moves. However, in
the limiting case that the strength of scattering is small, i.e.,
the probe is weakly interacting with the sample, STP measurement
result is not expected to change much. STP does have this
advantage of perturbing the sample minimally, compared to other
contact-based probes, for example, using conducting cantilevers as the fourth electrode and using an atomic force
microscope to scan on the sample. Conceptually, a good criteria for minimal source
effect would be that the tip should be far away from the sample
that it is in the far tail of the exponential decay of electron
wave function for sample states, though this will correspond to a large tunneling resistance and the resolution of the STP measurement in this case is expected to be low due to large Johnson noise. Nevertheless, it is a valid starting point to work under this assumption to simplify the problem. When the sample is small compared to its characteristic transport lengths, the effect of the tip is generally large and these effects are alreaday explicitly included in the formalism of works on this case\cite{buttiker1,buttiker2}.

\begin{acknowledgements}
We would like to thank Supriyo Datta %at Purdue University
and Kirk H. Bevan %at Oak Ridge National Laboratory
for a critical reading of the original draft of this manuscript and their valuable comments and suggestions. Support for this work came from the Air Force Office of Scientific Research. One of us (WW) further acknowledges the generous support of a Stanford Graduate Fellowship.
\end{acknowledgements}

\end{document}